\documentclass{article}
\usepackage[nonatbib,preprint]{neurips_2021}

\usepackage{microtype}
\usepackage{graphicx}
\usepackage{booktabs} 
\usepackage{amsmath}
\usepackage{amsthm}
\usepackage{amssymb}
\usepackage{url}
\usepackage{enumitem}
\usepackage[noend,linesnumbered,ruled,vlined]{algorithm2e}
\usepackage{bbm}
\usepackage{pgfplots}
\pgfplotsset{compat=1.7}
\usepackage{subcaption}
\usepackage{wrapfig}
\usepackage{mathtools}
\usepackage{hyperref}
\usepackage{comment}

\newtheorem{definition}{Definition}
\newtheorem{theorem}{Theorem}
\newtheorem{lemma}{Lemma}
\newtheorem{corollary}{Corollary}

\newcommand{\system}{{\sc F$^2$ed-Learning}\xspace}
\newcommand{\agg}{{FilterL2}\xspace}

\definecolor{darkgreen}{rgb}{0.0, 0.5, 0}

\newcommand{\F}{Fig.}

\newcommand{\sm}{Suppl. Material}
\renewcommand{\S}{Sec.}
\newcommand{\A}{Alg.}

\newcommand{\ignore}[1]{}

\title{Towards Bidirectional Protection in Federated Learning}

\author{%
  Lun Wang\thanks{The authors contribute equally to this paper.} \\
  University of California, Berkeley\\
  \texttt{wanglun@berkeley.edu} \\
  \And Qi Pang\footnotemark[1]\\
  Hong Kong University of Science and Technology\\
  \texttt{qpangaa@cse.ust.hk}
  \AND Shuai Wang\\
  Hong Kong University of Science and Technology\\
  \texttt{shuaiw@cse.ust.hk}
  \And Dawn Song\\
  University of California, Berkeley\\
  \texttt{dawnsong@cs.berkeley.edu}
}
\begin{document}
\maketitle


\author{Lun Wang}
\begin{abstract}

Prior efforts in enhancing federated learning (FL) security fall into two
categories. At one end of the spectrum, some work uses secure aggregation
techniques to hide the individual client's updates and only reveal the
aggregated global update to a malicious server that strives to infer the clients'
privacy from their updates. At the other end of the spectrum, some work uses
Byzantine-robust FL protocols to suppress the influence of malicious clients'
updates. We present a federated learning protocol \system, which, for the first
time, offers \textit{bidirectional defense} to simultaneously combat against the
malicious centralized server and Byzantine malicious clients.
To defend against Byzantine malicious clients, \system provides dimension-free
estimation error by employing and calibrating a well-studied robust mean
estimator \agg. \system also leverages secure aggregation to protect clients
from a malicious server.
One key challenge of \system is to address the incompatibility between \agg and
secure aggregation schemes.
Concretely, \agg has to check the individual updates from clients whereas secure
aggregation hides those updates from the malicious server.
To this end, we propose a practical and highly effective solution to split the
clients into \textit{shards}, where \system securely aggregates each shard's
update and launches \agg on updates from different shards.
The evaluation shows that \system consistently achieves optimal or
close-to-optimal performance and outperforms five secure FL protocols
under five popular attacks. 
%
%
\end{abstract}

\vspace{-0.2em}
\section{Introduction}
\label{sec:intro}
\vspace{-0.6em}

Federated learning (FL) has drawn numerous attention in the past few years as a
new distributed learning paradigm.
In FL, the users collaboratively train a model with the help of a centralized
server when all the data is held locally to preserve the users' privacy.
To defeat a malicious centralized server that can infer client's information
through their updates, FL protocols have been enhanced with secure aggregation
technique~\cite{bonawitz2017practical} which hides the individual local updates
and only reveals the aggregated global update.
The graceful balance between utility and privacy popularizes FL in a variety of
sensitive applications such as Google GBoard, healthcare service and
self-driving cars.

In addition to mitigating malicious servers, recent attacks have shown that a
small number of clients can behave maliciously in a large-scale FL system with
thousands of clients and stealthily influence the jointly-trained FL model.
In fact, for most SGD-based FL algorithms used
today~\cite{mcmahan2017federated}, the centralized server averages the local
updates to obtain the global update, which is vulnerable to even only one
malicious client. Particularly, a malicious client can arbitrarily craft its
update to either prevent the global model from converging or lead it to a
sub-optimal minimum.

To date, attacks over malicious clients have been
well-studied~\cite{bhagoji2019analyzing, fang2019local, bagdasaryan2020backdoor,
sun2020data}.
In particular, various Byzantine-robust FL
protocols~\cite{blanchard2017machine,yin2018byzantine,fu2019attack,pillutla2019robust}
are proposed to reduce the impact of the contaminated updates.
These protocols replace trivial averaging with well-designed Byzantine-robust
mean estimators, which suppress the influence of the malicious updates and
output a mean estimation as accurately as possible.
Nevertheless, these aggregators primarily suffer from the curse of
dimensionality.
Specifically, the estimation error scales up with the size of the model in a
square-root fashion.
As a concrete example, a three-layer MLP on MNIST contains more than 50,000
parameters and leads to a 223-fold increase of the estimation error, which is
prohibitive in practice.
Draco~\cite{chen2018draco}, BULYAN~\cite{mhamdi2018hidden} and
ByzatineSGD~\cite{alistarh2018byzantine} are the only three works that state to
yield dimension-free estimation error.
However, Draco is designed for distributed learning and is incompatible with
FL because it requires redundant updates from each worker.
Furthermore, while Bulyan~\cite{mhamdi2018hidden} and
ByzantineSGD~\cite{alistarh2018byzantine} provide dimension-free estimation
errors, they are based on much stronger assumptions than other contemporary
works.
As will be discussed in \S~\ref{sec:related_work}, when the assumptions are
relaxed to the common case, Bulyan's estimation error still scales up with the
square root of the model size.

Even worse, orchestrating robust FL estimators (to mitigate malicious clients)
with secure aggregation schemes (to mitigate malicious servers) is infeasible
for existing FL protocols: the robust estimators have to access local updates
whereas secure aggregation schemes generally hide them from the server.
Consequently, de facto FL protocols cannot \textit{simultaneously} protect the
server and the clients, but has to place complete trust in either of them.
The lack of two-way protection severely harms the dependability of FL systems
and generally prevents FL from being used in many real-world security-sensitive
applications such as home monitoring and automatic driving, where both servers
and clients could behave maliciously.

In this paper, we propose {\sc \textbf{Fed}erated \textbf{Learning}
with \textbf{F}ence}, abbreviately \system, a principled FL protocol to defend
against both the Byzantine malicious clients and the malicious server.
\system overcomes limitations of existing Byzantine-robust FL protocols
by employing and calibrating a well-established robust mean
estimator \agg~\cite{steinhardt2018robust} in FL scenarios. To address the
incompatibility issue between the robust mean estimator and secure
aggregation~\cite{bonawitz2017practical}, we propose a systematic and
highly-effective scheme to first split clients into \textit{shards}, where local
updates from the same shard are securely aggregated at the centralized server,
and the robust estimator is launched on the aggregated local updates from
different shards instead of individual clients.
Furthermore, we note that robust mean estimators like \agg\ were not used by previous works, given its strong assumption on i.i.d. updates from benign FL clients. Nevertheless, in this work, we show that when a shard contains reasonably large number of clients, it can practically aggregate non-i.i.d. updates into i.i.d..
That is, sharding enables novel opportunities to smoothly use robust mean estimators
like \agg in mitigating malicious clients.
We evaluate \system under five frequently-launched attacks over two datasets,
and compare \system with five robust estimators. Evaluation results show
that \system consistently achieves optimal or close-to-optimal performance under
all the attacks. We also studied how different settings of shards can influence
the security guarantees.
%
In summary, we make the following contributions:

\noindent {\tikz\draw[black,fill=black] (-0.5em,-0.5em) circle (-0.15em);} We
propose \system, the first FL protocol featuring principled and practical
defense \textit{simultaneously} against a malicious server and Byzantine
malicious clients. We propose the sharding scheme to reconcile dimension-free
robust estimators and secure aggregation. We also rigorously prove the
robustness and security guarantee of \system.
\vspace{-0.5em}

\noindent {\tikz\draw[black,fill=black] (-0.5em,-0.5em) circle (-0.15em);} We
point out the limitations in existing robust estimators with claimed
dimension-free errors. We reuse and calibrate a well-studied robust mean
estimation, \agg, to deliver a dimension-free estimation error in FL.
\vspace{-0.5em}

\noindent {\tikz\draw[black,fill=black] (-0.5em,-0.5em) circle (-0.15em);} Our
evaluation shows that \system can notably outperform existing robust estimators
in the presence of five popular attacks by always achieving optimal or close-to-optimal
performance.


\vspace{-0.8em}
\section{Related Work \& Limitations in Existing Byzantine-Robust Protocols}
\label{sec:related_work}
\vspace{-0.6em}

In this section, we review FL client privacy leakage and Byzantine malicious
client attacks: they are particularly addressed in this research. We also review
existing defenses and discuss their limitations that motivate this research.
For other known attacks and defenses in FL, we refer the interested readers to
the excellent surveys~\cite{kairouz2019advances,lyu2020threats}.

\vspace{-0.3em}
\noindent \textbf{Client Privacy Leakage and Mitigation.}~The inference attacks
in centralized learning~\cite{shokri2017membership,fredrikson2015model} aim to infer the private information of the model training data. Wang et
al.~\cite{wang2019beyond} explore the feasibility of recovering user privacy
from a malicious server in the collaborative (federated) learning settings. Nasr
et al.~\cite{nasr2019comprehensive} show that a malicious server can perform
highly accurate membership inference attacks against clients.
To enhance the privacy of clients, Bonawitz et al.~\cite{bonawitz2017practical}
propose secure aggregation, which provides security guarantee against the
malicious server. Also, by utilizing secure multi-party computation (MPC),
Mohassel et al.~\cite{mohassel2017secureml} present a framework where a global
model is trained on the clients' encrypted data among two non-colluding servers.
However, these existing defense methods generally assume benign clients, which
is not always realistic in practice.

\vspace{-0.3em}
\noindent \textbf{Byzantine Malicious Clients.}~Byzantine-robust aggregation has
drawn enormous attention in the past few years due to the emergence of various
distributed attacks in FL. Fang et al.~\cite{fang2019local} formalize the attack
as an optimization problem and successfully migrate the data poisoning attack to
FL. The proposed attacks even work under Byzantine-robust FL. Sun et
al.~\cite{sun2020data} manage to launch data poisoning attacks on the multi-task
FL framework. Bhagoji et al.~\cite{bhagoji2019analyzing} and Bagdasaryan et
al.~\cite{bagdasaryan2020backdoor} manage to insert backdoor functionalities
into the model via local model poisoning or local model replacement. Xie et
al.~\cite{xie2019dba} propose to split one backdoor into several parts and
insert it into the global model. Chen et al.~\cite{chen2020backdoor} and Zheng
et al.~\cite{zizhanzhenglearning} separately migrate backdoor attacks to
federated meta-learning and federated reinforcement learning. Meanwhile, Sun et
al.~\cite{sun2019can} show that norm clipping and ``weak'' differential privacy
mitigate backdoor attacks in FL without hurting the overall performance.
However, Wang et al.~\cite{wang2020attack} refute the claim and illustrate that
robustness to backdoors requires model robustness to adversarial examples, a
major open problem believed to be hard.

\vspace{-0.3em}
\noindent \textbf{Byzantine-Robust Protocols.}~A variety of Byzantine-robust FL
protocols are proposed to defend against these attacks.
Krum~\cite{blanchard2017machine} picks the subset of updates with enough close
neighbors and averages the subset. Yin et al.~\cite{yin2018byzantine} leverage
robust estimators like trimmed mean or median to achieve
order-optimal statistical error rate under strongly convex assumptions. Fung et
al.~\cite{fung2018mitigating} propose a similar robust estimator relying on a
robust secure aggregation oracle based on the geometric median. Yin et
al.~\cite{yin2019defending} propose to use robust mean estimators to defend
against saddle point attack. Pillutla et al.~\cite{pillutla2019robust} study
Sybil attacks in FL and propose a defense based on the diversity
of client updates. Ozdayi et al.~\cite{ozdayipreventing} design a defense for
backdoor attacks in FL by adjusting server-side learning rate.
Mhamdi et al.~\cite{mhamdi2018hidden} point out that Krum, trimmed mean and
median all suffer from $\mathcal{O}(\sqrt{d})$ ($d$ is the model size)
estimation error and propose a general framework Bulyan to reduce the error to
$\mathcal{O}(1)$.

\vspace{-0.3em}
\noindent \textbf{Limitations of Existing Robust Estimators.}~We point out that
the improvement of Bulyan actually comes from its stronger assumption. In
particular, Bulyan assumes that the expectation of the distance between two benign
updates is bounded by a constant $\sigma_1$, while Krum assumes that the
distance is bounded by $\sigma_2\sqrt{d}$. We can easily see that if
$\sigma_1=\sigma_2\sqrt{d}$, Bulyan falls back to the same order of estimation
error as Krum. The same loophole exists in the analysis of
ByzantineSGD~\cite{alistarh2018byzantine}. Consequently, there is no known FL
protocol equipped with dimension-free estimation error to mitigate Byzantine
adversaries.


\vspace{-0.8em}
\section{Problem Setup}
\vspace{-0.6em}

\label{sec:threat-model}
In this section, we review the general pipeline of FL and introduce the threat
model and defense goal. We use bold lower-case letters (\emph{e.g.}
\textbf{a},\textbf{b},\textbf{c}) to denote vectors, and $[n]$ to donate
$1\cdots n$.
%
%


\vspace{-0.3em}
\noindent \textbf{FL Pipeline.}~In an FL system, there is one server
$\mathcal{S}$ and $n$ clients $\mathcal{C}_i, i\in[n]$. Each client holds data
samples drawn from some unknown distribution $\mathcal{D}$.
Let $\ell(\textbf{w};\textbf{z})$ be the loss function on the model parameter
$\textbf{w}\in \mathbb{R}^d$ and a data sample $\textbf{z}$.
Let
$\mathcal{L}(\textbf{w})=\mathbb{E}_{\textbf{z}\sim\mathcal{D}}[\ell(\textbf{w};\textbf{z})]$
be the population loss function.
Our goal is to learn the model $\textbf{w}$ such that the population loss function is minimized: $\textbf{w}^{*}=\arg\min_{\textbf{w}\in\mathcal{W}}\mathcal{L}(\textbf{w})$.
%
To learn $\textbf{w}^*$, the whole system runs a $T$-round FL protocol.
Initially, the server stores a global model $\textbf{w}_0$.
In the $t^{th}$ round, $\mathcal{S}$ broadcasts the global model $\textbf{w}_{t-1}$ to the $m$ clients.
The clients then run the local optimizers (e.g., SGD, Adam, RMSprop), compute the difference $\textbf{g}^{(i)}_t$ between the optimized model and the global model, and upload the difference to $\mathcal{S}$.
In the $t^{th}$ round, $\mathcal{S}$ takes the average of the differences and updates the global model $\textbf{w}_t=\textbf{w}_{t-1}+\frac{1}{n}\sum_{i=1}^n \textbf{g}_t^{(i)}$.
%


\vspace{-0.3em}
\noindent \textbf{Threat Model \& Defense Goal.}~We assume that the centralized
server $\mathcal{S}$ can be malicious. The server can launch whatever attacks
such as inference attacks using legitimate updates from the clients as the only
inputs. However, the server cannot deviate from the protocol for the sake of
regulation or reputation pressure. This makes a malicious server highly
stealthy. As a convention in this line of
works~\cite{yang2019federated,bonawitz2017practical,mohassel2017secureml}, the
malicious server is referred to as a \textit{semi-honest} server in the rest of
this paper. We further assume that clients are $\epsilon$-Byzantine malicious,
meaning that at most $\epsilon n$ clients are malicious: they can arbitrarily
deviate from the protocol and tamper with their own updates for profitable or
even mischief purposes. We also clarify that there is no collusion between the
server and the clients. That is, the server cannot disguise as clients or hire
clients to launch colluded attacks.

The defense goal is two-fold. Firstly, we would like to achieve a dimension-free
error for the mean estimation in each round. Let $\boldsymbol{\mu}$ be the true
mean of the benign distribution and the output of a protocol with contaminated
inputs be $\hat{\boldsymbol{\mu}}$. The estimation error is defined by the
$\ell_2$ distance between the true mean and the estimation
$\|\hat{\boldsymbol{\mu}}-\boldsymbol{\mu}\|_2$. We also would like to minimize
the server's ability to infer sensitive information of the clients. Formally
speaking, we would like to hide the client's individual update in the aggregate
of multiple updates to guarantee client-side privacy.

\vspace{-0.8em}
\section{\texorpdfstring{\system}{F2ed-Learning}: Robust Privacy-Preserving FL}
\vspace{-0.6em}

\begin{figure*}[htbp]
\centering
\hspace*{-20pt}
\includegraphics[width=\textwidth]{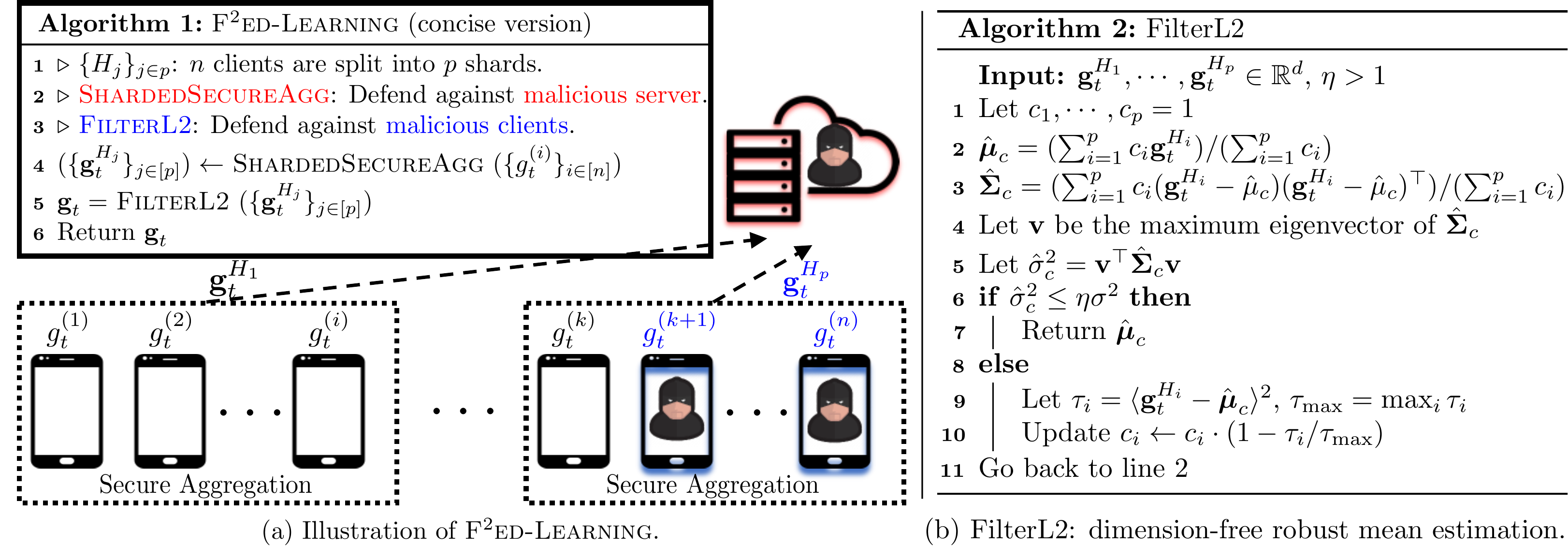}
\caption{High-level overview of \system and algorithm of \agg.}
\label{fig:pipeline}
\vspace{-10pt}
\end{figure*}

\label{sec:system}
The full protocol of \system will be given \A~\ref{alg:protocol}, and we first
summarize its high-level workflow in \F~\ref{fig:pipeline}(a): \system provides
bidirectional defense to simultaneously defend against the malicious server with
secure aggregation (marked in \textcolor{red}{red}) and malicious clients with
robust mean estimator \agg (marked in \textcolor{blue}{blue}).
\F~\ref{fig:pipeline}(b) formulates \agg, whose details are presented in
\S~\ref{sec:protocol1}. We now introduce \system step by step and formally
establish the robustness and security guarantees in \S~\ref{sec:analysis}. We
then discuss to what extent sharding can practically alleviate the i.i.d.
assumption required by \agg in \S~\ref{sec:iid}.

\captionsetup{font={footnotesize}}
%

\subsection{\texorpdfstring{\system}{F2ed-Learning}: Byzantine-Robust Privacy-Preserving FL}
\label{sec:protocol1}
The complete \system protocol is presented in Algorithm~\ref{alg:protocol}.
\system iteratively executes the following steps: (1) the server broadcasts the global model to the clients; (2) clients train the global model with their local data; (3) clients in the same shard run secure aggregation protocol to upload the mean of their updates to the server; (4) the server aggregates the received updates using robust mean estimation; (5) the server updates the global model with the aggregated global update.
We highlight steps (3) and (4) newly proposed in \system.
%

\vspace{-5pt}
\begin{algorithm}
\footnotesize
\DontPrintSemicolon
\SetKwBlock{Client}{Client:}{}
\SetKwBlock{Server}{Server:}{}
\SetKwInOut{Input}{Input}
\SetKwInOut{Output}{Output}
\SetKwProg{Fn}{Function}{:}{\KwRet}
\SetAlgoLined
\For{$t \leftarrow [T]$}{
    \Server{
        Split $n$ clients into $p$ shards $\{H_j\}_{j\in[p]}$\;
        
        Broadcast $\{H_j\}_{j\in [p]}$ and the global model $\textbf{w}_{t-1}$ to all the clients\;
    }
    \Client{
        \ForEach{client $i\in[n]$}{
            Locate its own shard $j$ and generate random masks $\textbf{u}^{(j)}_{ik}, k\in H_j/i$\;
            \lForEach{$k\in H_j/i$}{
                Send $\textbf{u}_{ik}$ to $k$
            }
            Train the local model $\textbf{w}_t^{(i)}$ using $\textbf{w}_{t-1}$ as initialization\;
            $\textbf{g}_t^{(i)}=\textbf{w}_t^{(i)} - \textbf{w}_{t-1} + \sum_{k\neq i,i\in H_j,k\in H_j}\textbf{u}_{ik}^{(j)}-\sum_{k\neq i,i\in H_j,k\in H_j}\textbf{u}_{ki}^{(j)}$\;
            Send $\textbf{g}_t^{(i)}$ to the server\;
        }
    }
    \Server{
        \lForEach{$H_j\in \{H_j\}_{j\in[p]}$} {
            $\textbf{g}_t^{H_j}=\frac{1}{|H_j|}\sum_{k\in H_j}\textbf{g}_t^{(k)}$
        }
        $\textbf{g}_t = \text{\agg}(\{\textbf{g}_t^{H_j}\}_{j\in[p]})$\;
        $\textbf{w}_t=\textbf{w}_{t-1}+\textbf{g}_t$\;
    }
}
\caption{\system: Robust Privacy-Preserving Sharded FL.}
\label{alg:protocol}
\end{algorithm}


\noindent \textbf{Sharded Secure Aggregation (lines 7--8, 10, 13).}~Secure
aggregation is developed by~\cite{bonawitz2017practical} to defend against a
semi-honest server in FL. Secure aggregation allows the server to obtain the sum
of the clients' updates but hides the individual updates cryptographically. We
use a simplified version of secure aggregation as follows for ease of
clarification. As the first step, each client samples random values for the
other clients and sends the values to the corresponding clients (lines 7--8).
After receiving all the values from other clients, each client sums up the
received values and subtracts the values generated by itself to produce a random
mask (line 10).
Each client blinds its local update with the random mask and sends the blinded
update to the server (line 11).
The server then sums up all the blinded updates and obtains the summed update in plaintext (line 13).
Obviously, all the masks cancel out during aggregation and the server receives the plaintext sum.
Secure aggregation provides strong privacy guarantee for the clients that the server cannot see anything but the aggregated global update and each client is hidden in thousands of other clients.

However, in our threat model, vanilla secure aggregation is insufficient since it provides no protection for the server.
As the individual updates are completely hidden, the server can not identify the malicious clients even after detecting the attack.
To address the issue, we propose to split the clients into multiple shards and run secure aggregation within each shard.
The size of the shards provides a trade-off between the protection for the
server and the clients.
The smaller the size is, the more information is revealed to the server, thus
becoming easier to defend against Byzantine malicious clients and harder to
fight off the semi-honest server.
The trade-off is discussed in detail in \S~\ref{sec:analysis}.

\vspace{-0.3em}
\noindent \textbf{Robust Mean Estimation (line 14).}~The core step in
Byzantine-robust FL is to estimate the true mean of the benign updates as
accurately as possible even with some malicious clients. The most commonly used
aggregator, averaging, is proven to be vulnerable to even only one malicious
client. Existing works in this field (e.g., Krum~\cite{blanchard2017machine} and
Bulyan~\cite{mhamdi2018hidden}), however, suffer from a
\textit{dimension-dependent} estimation error. Note that such
dimension-dependent error can impose intolerably high cost even for training a
3-layer MLP on MNIST, not to mention more complicated tasks and models such as
VGG16 or ResNet50.

It is worth mentioning that in statistics, robust mean estimation has been
well-studied to deliver a robust mean estimator with dimension-free estimation
error~\cite{diakonikolas2019robust, charikar2017learning, steinhardt2018robust,
  cheng2019high, dong2019quantum}. However, such dimension-free mean estimators
were not adopted, given their strict assumption on i.i.d. data. 

Soon in \S~\ref{sec:iid}, we prove that the sharding scheme can practically
convert individual clients' non-i.i.d. updates into an ideally i.i.d. shared update, 
suppose a shard contains a reasonably large amount of clients. This novelly
illustrates the feasibility to smoothly adopt dimension-free robust mean
estimators. Particularly, \system incorporates a well-known robust mean
estimator: \agg~\cite{steinhardt2018robust}. We formulate \agg in
\A~\hyperref[fig:pipeline]{2}. Specifically, \agg assigns each update a weight
and iteratively updates the weights until the weights for the malicious updates
are small enough. As mentioned, \agg provides a dimension-free error rate which is
formally presented as follows.

\begin{theorem}[\cite{steinhardt2018robust}]
Let $D$ be the honest dataset and $D^{*}$ be the contaminated version of $D$ by inserting malicious samples.
Suppose that $|D^{*}|\leq |D|/(1-\epsilon), \epsilon\leq\frac{1}{12}$, and further suppose that $\textsc{Mean}[D]=\boldsymbol{\mu}$ and $\|\textsc{Cov}[D]\|_{op}\leq\sigma^2$. %
Then given $\mathcal{D}^*$, Algorithm~\hyperref[fig:pipeline]{2} outputs $\hat{\boldsymbol{\mu}}$ s.t. $\|\hat{\boldsymbol{\mu}}-\boldsymbol{\mu}\|_2=\mathcal{O}(\sigma\sqrt{\epsilon})$ using $\textsc{poly}(n,d)$ time.
%
%
\label{thm:filterl2}
\end{theorem}

Although Algorithm~\hyperref[fig:pipeline]{2} only takes polynomial time to run, the per-round time complexity is $\mathcal{O}(nd^2)$ if implemented with power iteration.
Given $d$ is large, the running time is still quite expensive in practice.
To address the issue, we cut the update vectors into $k$ sections and apply the robust estimator to each of the sections.
The acceleration scheme reduces the per-round running time to $\mathcal{O}(nd^2/k)$ but increases the estimation error to $\mathcal{O}(\sigma\sqrt{\epsilon k})$.
For instance, if we take $k=\sqrt{d}$, the per-round running time becomes
$\mathcal{O}(nd)$ whereas the estimation error grows to
$\mathcal{O}(\sigma\sqrt[4]{\epsilon^2 d})$.
Despite the tradeoff for acceleration, \agg still gives the known optimal
estimation error and outperforms other robust FL protocols by multiple
magnitudes, as will be shown in \S~\ref{sec:evaluation}.

\vspace{-0.8em}
\subsection{Robustness \& Security Analysis}
\label{sec:analysis}
\vspace{-0.6em}

In this section, we rigorously prove the security and robustness guarantee of \system.

\vspace{-0.3em}
\noindent \textbf{Security Guarantee.}~We first give the security guarantee of
\system as follows. Intuitively, no more information about the clients except
the averaged updates from the shards is revealed to the centralized server.
Thus, each client's update is hidden by the rest clients in its shard.
%
%
\begin{corollary}[Security against semi-honest server; Informal]
Let $\Pi$ be an instantiation of \system, there exists a PPT (probabilistic polynomial Turing machine) simulator $\textsc{Sim}$ which can only see the averaged updates from the shards. For all clients $\mathcal{C}$, the output of $\textsc{Sim}$ is computationally indistinguishable from the view of that real server $\Pi_\mathcal{C}$ in that execution, i.e., $\Pi_\mathcal{C} \approx \textsc{Sim}(\{ \textbf{g}_t^{H_j} \}_{j \in [p]})$.

\label{cor:sec}

\end{corollary}

\vspace{-0.3em}
\noindent \textbf{Robustness Guarantee.}~We now give the formal robustness
guarantee of \system. The proof involves a trivial application of
Theorem~\ref{thm:filterl2} so we omit it here.
%
%

\begin{corollary}[Robustness against Byzantine adversaries]
Given the number of clients $n$, the number of shards $p$, and the fraction of corrupted clients $\epsilon$, \system provides a mean estimation with dimension-free error as long as $12\epsilon n < p$.
\label{cor:robust}
\end{corollary}





\vspace{-0.3em}
\noindent \textbf{Remark.}~Given the formal security and robustness guarantee,
we can see that \system provides a convenient way to calibrate the
protection for the server or the clients.
Concretely, \system can tolerate up to $\lfloor\frac{p}{12}\rfloor-1$ malicious clients and hide each honest client's update in the mean of $\lfloor\frac{n}{p}\rfloor$ updates.

The full proofs of Corollary~\ref{cor:sec} and Corollary~\ref{cor:robust} are
given in Appendix~\ref{sec:proof-corollary1} and
Appendix~\ref{sec:proof-corollary2}, respectively. Note that the proof of
Corollary~\ref{cor:robust} assumes the updates from the benign shards are
i.i.d.; the following section justifies this assumption.

\vspace{-0.8em}
\subsection{Discussion on the i.i.d. Assumption in Corollary~\ref{cor:robust}}
\label{sec:iid}
\vspace{-0.6em}

To derive Corollary~\ref{cor:robust}, we assume that the updates from the benign
shards are drawn from some i.i.d. distribution $\mathcal{D}$. In this section,
we explore the rationality of the assumption.
In general, if the benign updates are drawn from distributions that diverse
largely, the Byzantine-robust estimators are difficult to identify and rule out
malicious updates. Therefore, robust mean estimators like \agg, though offering
dimension-free error estimation, were not commonly used in this field given
their requirement on i.i.d. data.
However, we show that when given reasonably large shard size, non-i.i.d. data
becomes i.i.d. after sharding; we present corresponding proofs in
Corollary~\ref{cor:clt} and empirical results in \S~\ref{sec:eval-results}. As a
result, sharding novelly enables the adoption of FilterL2 in the FL scenarios.
On the other hand, \textit{we clarify that we are not addressing the accuracy
  drop caused by non-i.i.d. data in FL. Instead, our discussion in this section
  and empirical results in \S~\ref{sec:eval-results} validate the usage of
  FilterL2 since the dimension-free guarantee generally holds under the i.i.d.
  assumption.}

\vspace{-0.3em}
\noindent \textbf{Two Sources of Non-i.i.d. Updates in FL.}~It is well known
that in FL, data is heterogeneously distributed across clients. Therefore, the
collected updates are typically not i.i.d. under any proper distribution.
Another source of non-i.i.d. updates in FL is the random initialization of local
models. As known, many neural networks are permutation-invariant. For instance,
in a two-layer fully connected network, the neurons in the two layers can be
permuted correspondingly without changing the functionality of the network.
Therefore, even with the same training data, different initialization can lead
to different models within the same permutation-invariant class.

Hence, to overcome the second source of non-i.i.d. updates, we take the standard
approach by requiring the clients to share the same initialization before the
training phase starts. Note that there is a line of
works~\cite{yurochkin2019statistical,yurochkin2019bayesian,wang2020federated}
focusing on automatically addressing this issue using matching algorithms and
Bayesian non-parametric models. We deem it as an interesting future direction to
incorporate these works in \system.

For the rest of the section, we discuss to what extent \system alleviates the
first source of non-i.i.d. updates. We formally model the heterogeneous data
distribution under explicit assumptions and discuss how sharding addresses the
first issue under such assumptions. Note that with sharding we do not solve the
slow convergence issue in FL due to non-i.i.d. updates. Instead, we only create
a distribution that is i.i.d. among shards given the shard size is reasonably
large to practically satisfy the i.i.d. assumption required in the proof of
Corollary~\ref{cor:robust}.
The distribution is highly biased and still suffers from a low convergence rate
and accuracy drop due to the intrinsic non-i.i.d. data.
%

%


%
Now we propose a novel perspective to conduct robustness analysis in FL.
Succinctly, by aggregating the shards first, we can reduce the
non-i.i.d. updates to i.i.d. when the shard size is reasonably large, given some
assumptions on the non-i.i.d. updates. As the first step, we introduce the
assumption on the non-i.i.d. distribution in Definition~\ref{def:hetero}, whose
validity stems from the observation that a major source of
heterogeneity in classification task is the unbalanced distribution of data with
different labels~\cite{mcmahan2017communication,li2018federated}. 

\begin{definition}[Heterogeneous Distribution]
Let $\mathbb{D}$ be a set of $k$ distributions $\mathbb{D}=\{\mathcal{D}_i\}_{i\in[k]}$ where $\mathbb{E}[\mathcal{D}_i]=\mu_i$ and $\mathbb{V}[\mathcal{D}_i]=\sigma^2_i$.
Each client $\mathcal{C}_j$'s update $\textbf{g}_j$ follows a distribution $\mathcal{D}_{\phi(j)}$ where $\phi$ is a mapping from the client index to the distribution index.
\label{def:hetero}
\end{definition}

%

As the second step, we analyze the influence of sharding on the update distribution.
Ideally, when the shard size is reasonably large, sharding pushes the non-i.i.d. distribution to a well-regulated i.i.d. distribution according to Theorem~\ref{thm:lindeberg}.
\begin{theorem}[Lindeberg Central Limit Theorem~(\cite{linnik1959information})]
Suppose $\{X_1, \cdots, X_n\}$ is a sequence of independent random variables (not necessarily identically distributed), each with finite expected value $\mu_i$ and variance $\sigma_i^2$.
Define $s_n^2=\sum_{i=1}^n\sigma_i^2$.
Suppose that $\forall \epsilon > 0$,
\begin{equation*}
    \lim_{n\rightarrow \infty}\frac{1}{s_n^2}\sum_{i=1}^n\mathbb{E}[(X_i-\mu_i)^2\cdot\mathbbm{1}\{|X_i-\mu_i|>\epsilon s_n\}]=0.
\end{equation*}
Then, the distribution of the standardized sums converges towards the standard normal distribution.
\begin{equation}
\frac{1}{s_n}\sum_{i=1}^{n}(X_i-\mu_i)\overset{d}{\rightarrow} N(0,1)
\label{eq:clt}
\end{equation}

\label{thm:lindeberg}
\end{theorem}
\vspace{-5pt}
Given Definition~\ref{def:hetero} and Theorem~\ref{thm:lindeberg}, the following corollary follows naturally.

\begin{corollary}[Ideally i.i.d. after sharding]
Assume that the updates from the clients follow Definition~\ref{def:hetero} where $k \ll \frac{n}{p}$.
Besides, 
\vspace{-5pt}
\begin{equation*}
\lim_{|H|\rightarrow \infty}\frac{1}{s_H^2}\sum_{i\in H}\mathbb{E}[(g_i-\mu_i)^2\cdot\mathbbm{1}\{|g_i-\mu_i|>\epsilon s_H\}]=0
\end{equation*}
, where $s_H^2=\sum_{i\in H}\sigma_{\phi(i)}^2$.
Given the uniform randomness of sharding, we can view the distribution index $\phi(j)$ as drawn from some distribution $\Phi$ on $[k]$.
Let $\bar{\mu}=\mathbb{E}_{i\sim\Phi}[\mu_i]=\sum_{x\in[k]}\Phi(x=i)\mu_i$ and $\bar{\sigma}^2=\mathbb{E}_{i\sim\Phi}[\sigma_i^2]=\sum_{x\in[k]}\Phi(x=i)\sigma_i^2$.
Then, 
\vspace{-5pt}
\begin{equation*}
    \frac{1}{|H|}\sum_{i\in H}g_i\overset{d}{\rightarrow} N(\bar{\mu}, \frac{\bar{\sigma}^2}{|H|})
\end{equation*}

\label{cor:clt}
\end{corollary}
\vspace{-5pt}
We provide the proof of Corollary~\ref{cor:clt} in Appendix~\ref{sec:proof-corollary3}.
Besides, an empirical validation is also reported in Appendix~\ref{sec:emp-val}, 
which illustrates that the shards' updates are more densely and identically
distributed compared with the non-i.i.d. individual updates.

\vspace{-0.8em}
\section{Evaluation}
\label{sec:evaluation}
\vspace{-0.6em}

In this section, we want to answer the following questions using empirical
evaluation: (\textbf{$Q_1$}) Does \agg outperform other aggregators when used
alone? (\textbf{$Q_2$}) Does \system outperform other robust FL protocols
augmented with sharded secure aggregation? (\textbf{$Q_3$}) How does the shard
size affect the performance of \system?

%
\vspace{-0.3em}
\noindent \textbf{Attacks.}~We evaluated the robust estimators without attack
and with several representative attacks via malicious clients which are Krum
Attack (KA)~\cite{fang2019local}, Trimmed Mean Attack
(TMA)~\cite{fang2019local}, Model Poisoning Attack
(MPA)~\cite{bhagoji2019analyzing}, Model Replacement Attack
(MRA)~\cite{bagdasaryan2020backdoor}, and Distributed Backdoor Attack
(DBA)~\cite{xie2019dba}. Please refer to Appendix~\ref{sec:eval-attack} for the
details of these attacks.

\vspace{-0.3em}
\noindent \textbf{Experimental Setup.}~We selected two datasets,
MNIST~\cite{lecun2010mnist} and FashionMNIST~\cite{xiao2017fashion}, to evaluate
\system. We also chose three other Byzantine-robust FL protocols as baselines:
(1) Krum~\cite{blanchard2017machine}; (2) Trimmed Mean~\cite{yin2018byzantine};
and (3) Bulyan~\cite{mhamdi2018hidden}. Note that Bulyan acts like a wrapper
around other robust estimators. Therefore, in the evaluation, we have two
versions of Bulyan: Bulyan Krum and Bulyan Trimmed Mean. We run all the
protocols on the two datasets and present the protocols' performance under
different attacks. Performance is measured differently according to 
different attack targets. For KA and TMA, we use the model accuracy as the
metric for characterizing attack performance. Higher model accuracy indicates
stronger robustness. For MPA, MRA, and DBA, we assess the percentage of the
remembered backdoors to demonstrate the attack performance. The fewer backdoors
remembered, the more robust the estimator is.

We evaluate \system under both homogeneous (i.i.d.) and heterogeneous
(non-i.i.d.) data distributions. In the i.i.d. setting, the data is randomly
partitioned into 20 clients, each receiving the same number of examples. Five
out of the 20 clients are malicious.
In the non-i.i.d. setting, we set up 100 clients, ten of which are malicious, and
each client is assigned data with three labels. The clients are randomly split
into 25 shards. For other details like model architecture and hyper-parameters,
please refer to Appendix~\ref{sec:eval-setup-append}.
%

\newcommand{\figwidth}{0.24\textwidth}
\newcommand{\lfigwidth}{0.43\textwidth}
\tikzset{font={\fontsize{15pt}{12}\selectfont}}
\captionsetup{font={footnotesize},skip=2pt}
\captionsetup[sub]{font={footnotesize},skip=2pt}
\begin{figure*}[htbp]
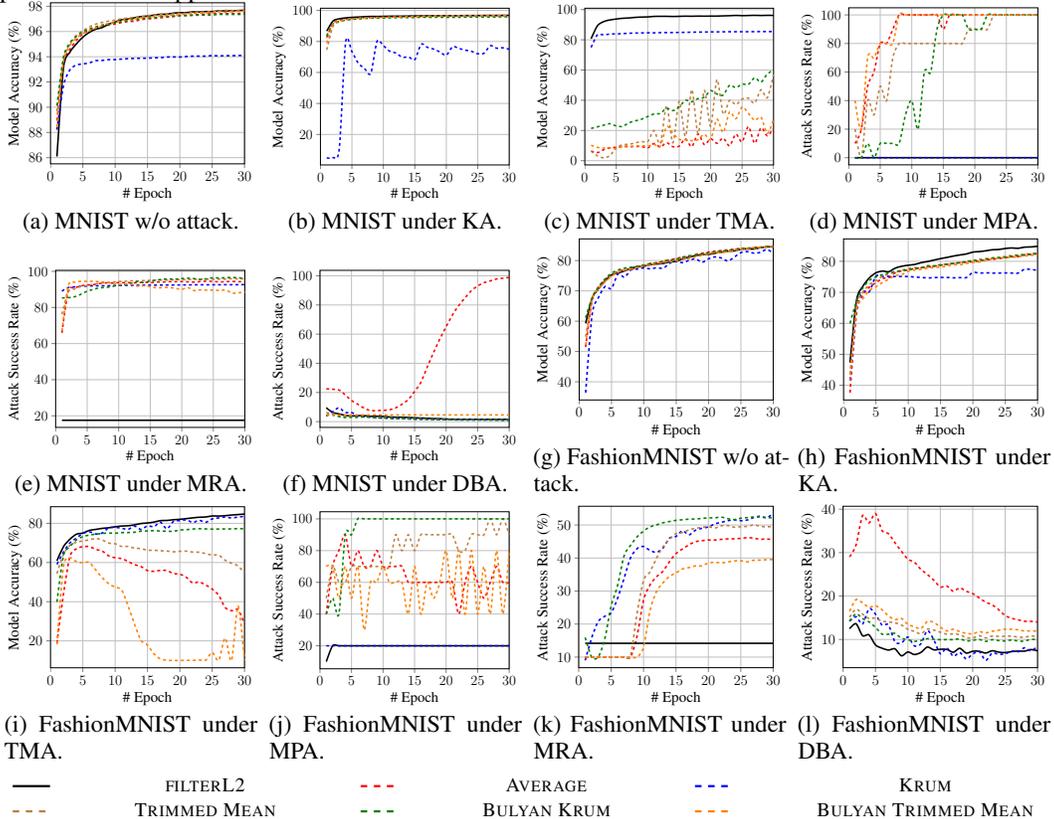

    \vspace{-15pt}
     \centering
     \hfill
     \begin{subfigure}[b]{\figwidth}
         \centering
         \resizebox{\textwidth}{!}{\input{figs/main/orig/INFIMNIST/withoutattack}}
         \caption{MNIST w/o attack.}
         \label{fig:mnist_noattack}
     \end{subfigure}
     \hfill
     \begin{subfigure}[b]{\figwidth}
         \centering
         \resizebox{\textwidth}{!}{\input{figs/main/orig/INFIMNIST/krum}}
         \caption{MNIST under KA.}
         \label{fig:mnist_krum}
     \end{subfigure}
     \hfill
     \begin{subfigure}[b]{\figwidth}
         \centering
         \resizebox{\textwidth}{!}{\input{figs/main/orig/INFIMNIST/trimmedmean}}
         \caption{MNIST under TMA.}
         \label{fig:mnist_trimmedmean}
     \end{subfigure}
     \hfill
     \begin{subfigure}[b]{\figwidth}
         \centering
         \resizebox{\textwidth}{!}{\input{figs/main/orig/INFIMNIST/modelpoisoning}}
         \caption{MNIST under MPA.}
         \label{fig:mnist_modelpoisoning}
     \end{subfigure}

     \hfill
    \begin{subfigure}[b]{\figwidth}
        \centering
        \resizebox{\textwidth}{!}{\input{figs/main/orig/INFIMNIST/backdoor}}
        \caption{MNIST under MRA.}
        \label{fig:mnist_backdoor}
    \end{subfigure}
     \hfill
    \begin{subfigure}[b]{\figwidth}
        \centering
        \resizebox{\textwidth}{!}{\input{figs/main/orig/INFIMNIST/dba}}
        \caption{MNIST under DBA.}
        \label{fig:mnist_dba}
    \end{subfigure}
    \hfill
     \begin{subfigure}[b]{\figwidth}
         \centering
         \resizebox{\textwidth}{!}{\input{figs/main/orig/fashionMNIST/withoutattack}}
         \caption{FashionMNIST w/o attack.}
         \label{fig:fasionmnist_noattack}
     \end{subfigure}
     \hfill 
     \begin{subfigure}[b]{\figwidth}
         \centering
         \resizebox{\textwidth}{!}{\input{figs/main/orig/fashionMNIST/krum}}
         \caption{FashionMNIST under KA.}
         \label{fig:fasionmnist_krum}
     \end{subfigure}
     
     \hfill
     \begin{subfigure}[b]{\figwidth}
         \centering
         \resizebox{\textwidth}{!}{\input{figs/main/orig/fashionMNIST/trimmedmean}}
         \caption{FashionMNIST under TMA.}
         \label{fig:fashionmnist_trimmedmean}
     \end{subfigure}
     \hfill
     \begin{subfigure}[b]{\figwidth}
         \centering
         \resizebox{\textwidth}{!}{\input{figs/main/orig/fashionMNIST/modelpoisoning}}
         \caption{FashionMNIST under MPA.}
         \label{fig:fashionmnist_modelpoisoning}
     \end{subfigure}
    \hfill
    \begin{subfigure}[b]{\figwidth}
        \resizebox{\textwidth}{!}{\input{figs/main/orig/fashionMNIST/backdoor}}
        \caption{FashionMNIST under MRA.}
        \label{fig:fashionmnist_backdoor}
    \end{subfigure}
    \hfill
    \begin{subfigure}[b]{\figwidth}
        \resizebox{\textwidth}{!}{\input{figs/main/orig/fashionMNIST/dba}}
        \caption{FashionMNIST under DBA.}
        \label{fig:fashionmnist_dba}
    \end{subfigure}

    \begin{subfigure}[t]{\textwidth}
        \centering
        \resizebox{\textwidth}{!}{\input{figs/main/legend}}
    \end{subfigure}
    \caption{Attack performance under various Byzantine-robust estimators with i.i.d. data.}
    \label{fig:f1}
    \vspace{-15pt}
\end{figure*}

\vspace{-0.8em}
\subsection{Evaluation Results}
\label{sec:eval-results}
\vspace{-0.6em}

In this section, we present the evaluation results.
We first show that \agg outperforms other robust aggregators when used alone.
Then, we run complete \system with sharding and the results show that \system consistently achieves optimal or close-to-optimal performance under different attacks.
Last, we discuss the effect of the shard size on the performance.

\noindent\textbf{\agg Performance without Sharding.}
To answer question \textbf{$Q_1$}, we evaluated six aggregators on MNIST and
FashionMNIST under the i.i.d. setting as shown in \F~\ref{fig:f1}. Besides, we
report the corresponding evaluation result on MNIST under the non-i.i.d. setting
in \F~\ref{fig:f4} of Appendix~\ref{sec:eval-non-iid}.
%
%
%


In the i.i.d. setting, \agg achieves optimal performance among all 6 aggregators.
For MNIST, under KA and TMA, \agg separately achieves 96.85\% and 96.15\% accuracy within 30 epochs, comparable to the non-malicious setting with accuracy 97.48\%.
Under MPA, MRA, and DBA, \agg separately reduces the attack success rate to 0.00\%, 17.53\% and 1.62\%.
For Fashion-MNIST, under KA and TMA, \agg separately achieves 84.87\% and 84.75\% accuracy, slightly better than the non-malicious setting with accuracy 84.66\%.
Under MPA, MRA, and DBA, \agg manages to suppress the attack success rate to 20.00\%, 14.18\%, and 7.38\%.
Specifically, \agg is the only aggregator that consistently achieves good performance under all five attacks.

In the non-i.i.d. setting, most of the estimators do not perform well as the
i.i.d. assumption is broken (without splitting clients into shards). Notably,
\agg still achieves 91.75\% accuracy under TMA and reduces the attack success
rate to 21.46\% and 0.86\% at the end of the training process under MRA and DBA.
However, the accuracy drops to 81.01\% under KA, and the attack success rate
increases to 100.00\% under MPA. This is intuitive, given that the i.i.d.
assumption on which \agg relies is broken.

\vspace{-0.3em}
\noindent \textbf{\textsc{F$^2$ed-Learning} Performance.}~To answer question \textbf{$Q_2$}, we
evaluate six aggregators with sharding on MNIST and FashionMNIST under the
i.i.d. setting as shown in \F~\ref{fig:f2}. Besides, the corresponding result of
MNIST under the non-i.i.d. setting is shown in \F~\ref{fig:f5} of
Appendix~\ref{sec:eval-non-iid}. We run the protocols with 100 clients, ten of
which are malicious.
%
%
%

\tikzset{font={\fontsize{15pt}{12}\selectfont}}
\captionsetup{font={footnotesize},skip=2pt}
\captionsetup[sub]{font={footnotesize},skip=2pt}
\begin{figure*}[htbp]
\vspace{-15pt}
     \centering
     \hfill
     \begin{subfigure}[b]{\figwidth}
         \centering
         \resizebox{\textwidth}{!}{\input{figs/main/shard/INFIMNIST/withoutattack}}
         \caption{MNIST w/o attack.}
         \label{fig:mnist_noattack2}
     \end{subfigure}
     \hfill
     \begin{subfigure}[b]{\figwidth}
         \centering
         \resizebox{\textwidth}{!}{\input{figs/main/shard/INFIMNIST/krum}}
         \caption{MNIST under KA.}
         \label{fig:mnist_krum2}
     \end{subfigure}
     \hfill
     \begin{subfigure}[b]{\figwidth}
         \centering
         \resizebox{\textwidth}{!}{\input{figs/main/shard/INFIMNIST/trimmedmean}}
         \caption{MNIST under TMA.}
         \label{fig:mnist_trimmedmean2}
     \end{subfigure}
     \hfill
     \begin{subfigure}[b]{\figwidth}
         \centering
         \resizebox{\textwidth}{!}{\input{figs/main/shard/INFIMNIST/modelpoisoning}}
         \caption{MNIST under MPA.}
         \label{fig:mnist_modelpoisoning2}
     \end{subfigure}
     
     \hfill
     \begin{subfigure}[b]{\figwidth}
         \centering
         \resizebox{\textwidth}{!}{\input{figs/main/shard/INFIMNIST/backdoor}}
         \caption{MNIST under MRA.}
         \label{fig:mnist_backdoor2}
     \end{subfigure}
     \hfill
     \begin{subfigure}[b]{\figwidth}
         \centering
         \resizebox{\textwidth}{!}{\input{figs/main/shard/INFIMNIST/dba}}
         \caption{MNIST under DBA.}
         \label{fig:mnist_dba2}
     \end{subfigure}
     \hfill     
     \begin{subfigure}[b]{\figwidth}
         \centering
         \resizebox{\textwidth}{!}{\input{figs/main/shard/fashionMNIST/withoutattack}}
         \caption{FashionMNIST w/o attack.}
         \label{fig:fashionmnist_noattack2}
     \end{subfigure}
     \hfill
     \begin{subfigure}[b]{\figwidth}
         \centering
         \resizebox{\textwidth}{!}{\input{figs/main/shard/fashionMNIST/krum}}
         \caption{FashionMNIST under KA.}
         \label{fig:fashionmnist_krum2}
     \end{subfigure}
     
     \hfill
     \begin{subfigure}[b]{\figwidth}
         \centering
         \resizebox{\textwidth}{!}{\input{figs/main/shard/fashionMNIST/trimmedmean}}
         \caption{FashionMNIST under TMA.}
         \label{fig:fashionmnist_trimmedmean2}
     \end{subfigure}
     \hfill
     \begin{subfigure}[b]{\figwidth}
         \centering
         \resizebox{\textwidth}{!}{\input{figs/main/shard/fashionMNIST/modelpoisoning}}
         \caption{FashionMNIST under MPA.}
         \label{fig:fashionmnist_modelpoisoning2}
     \end{subfigure}
     \hfill
     \hfill
     \begin{subfigure}[b]{\figwidth}
         \centering
         \resizebox{\textwidth}{!}{\input{figs/main/shard/fashionMNIST/backdoor}}
         \caption{FashionMNIST under MRA.}
         \label{fig:fashionmnist_backdoor2}
     \end{subfigure}
     \hfill
     \begin{subfigure}[b]{\figwidth}
         \centering
         \resizebox{\textwidth}{!}{\input{figs/main/shard/fashionMNIST/dba}}
         \caption{FashionMNIST under DBA.}
         \label{fig:fashionmnist_dba2}
     \end{subfigure}
     
     
    \begin{subfigure}[t]{\textwidth}
        \centering
        \resizebox{\textwidth}{!}{\input{figs/main/legend2}}
    \end{subfigure}
    \caption{Attack performance under various estimators with sharded secure aggregation and i.i.d. data.}
    \label{fig:f2}
    \vspace{-18pt}
\end{figure*}

In both i.i.d. and non-i.i.d. settings, for the experiments without attack, with TMA or with MPA~(\F~\ref{fig:mnist_noattack2},\ref{fig:mnist_trimmedmean2},\ref{fig:mnist_modelpoisoning2},\ref{fig:fashionmnist_noattack2}.\ref{fig:fashionmnist_trimmedmean2},\ref{fig:fashionmnist_modelpoisoning2},\ref{fig:mnist_noattack4},\ref{fig:mnist_trimmedmean4},\ref{fig:mnist_modelpoisoning4}), \system still achieves optimal or close-to-optimal performance.
In the i.i.d. setting, \system achieves 94.86\% and 95.38\% accuracy under KA
and TMA on MNIST while averaging with sharding achieves 95.41\% without attack.
It also suppresses the attack success rate to 0.00\%, 10.49\%, and 3.52\% under MPA, MRA, and DBA.
On FashionMNIST, \system achieves 83.22\% and 84.12\% under KA and TMA while the baseline is 83.87\% without attack.
Under MPA, MRA, and DBA, the attack success rate is controlled under 20.00\%, 14.41\%, and 2.78\% at the end of training.

Moreover, in the non-i.i.d. setting, \system achieves 92.96\% accuracy,
improving 11.95\% compared with \agg without sharding. \system also reduces the
attack success rate under MPA to 40.00\% compared with 100.00\% without sharding.
This improvement is due to the fact that sharding re-establishes the i.i.d.
nature of the collected updates --- the necessity for \agg to be functional as
discussed in \S~\ref{sec:iid}.



%
%

\vspace{-0.3em}
\noindent \textbf{Influence of Shard Size.}~To answer question \textbf{$Q_3$},
we empirically evaluate the influence of shards number $p$.
Generally, with a proper $p$, \system will provide both security and robustness guarantee.
However, too large $p$ will weaken its security guarantee and too small $p$ will weaken its robustness.
Please refer to Appendix~\ref{sec:eval-shardsize} for the detailed evaluation results and illustration.
%

\vspace{-0.8em}
\section{Limitation \& Conclusion}
\vspace{-0.6em}
\label{sec:conclusion}

In this paper, we designed and developed \system, the first FL protocol defending against a semi-honest server and Byzantine malicious clients simultaneously.
We propose to use \agg to robustly aggregate the possibly contaminated updates and secure aggregation to protect the privacy of the clients.
We reconcile the contradictory components with sharding.
The evaluation results show that \system consistently achieves the optimal or close-to-optimal performance among five robust FL protocols.
As far as we can see, \system addresses the two main privacy threats in FL systems simultaneously and shows
the potential to further popularize FL in sensitive applications.

We also identify several unsolved challenges in \system which might motivate future works in FL with bidirectional protection.
For instance, vanilla \agg brings extra overhead due to its nearly-quadratic complexity.
Although the accelerated \agg partially addresses the issue, it sacrifices the asymptotic estimation error for the speedup.
An interesting future direction is to integrate robust mean estimators with low complexity such as~\cite{cheng2019high}.
However, Cheng et al.'s approach proposed in~\cite{cheng2019high} is rather complicated so designing a low-complexity robust mean estimator with simple intuition is also an intriguing direction.
%


\bibliographystyle{plain}
\bibliography{ref.bib}


\section*{Appendix}
\appendix
\section{Proof of Corollary~\ref{cor:sec}}
\label{sec:proof-corollary1}

\noindent \textbf{Corollary~\ref{cor:sec}}~(Security against a semi-honest server; Informal).
Let $\Pi$ be an instantiation of \system, there exists a PPT (probabilistic polynomial Turing machine) simulator $\textsc{Sim}$ which can only see the averaged updates from the shards. For all clients $\mathcal{C}$, the output of $\textsc{Sim}$ is computationally indistinguishable from the view of that real server $\Pi_\mathcal{C}$ in that execution, i.e., $\Pi_\mathcal{C} \approx \textsc{Sim}(\{ \textbf{g}_t^{H_j} \}_{j \in [p]})$.

\begin{proof}
The transcript of the server is the updates from the sharded clients $\{ \textbf{g}_t^{H_j} \}_{j \in [p]}$.
Hence, Corollary~\ref{cor:sec} is equivalent to the following lemma since the $\textsc{Sim}$ can split the aggregated updates into several random shards which is computationally indistinguishable from the true transcript. 

\begin{lemma}[Lemma 6.1 in~\cite{bonawitz2017practical}]
Given any shard $H_k$ which is formed by a set of clients $C_k$, the parameter size $d$, the group size $q$, and the updates $\textbf{g}^{(i)}$ where $\forall i\in C_k$, $\textbf{g}^{(i)}\in \mathbb{Z}^d_q$, we have 
\begin{equation*}
\begin{split}
    &\{\{\textbf{u}_{ij}\overset{\$}{\leftarrow}\mathbb{Z}_q^d\}_{i<j}, \textbf{u}_{ij}\coloneqq -\textbf{u}_{ji} \; \forall \; i,j \in C_k, \; i > j \quad : \{\textbf{g}^{(i)}+\sum_{j\in C_k \slash i}\textbf{u}_{ij} \pmod{q}\}_{i\in C_k}\}
    \\ \equiv &\{\{\textbf{v}_i\overset{\$}{\leftarrow}\mathbb{Z}_q^d\}_{i\in C_k}~s.t.~ \sum_{i\in C_k}\textbf{v}_i=\sum_{i\in C_k}\textbf{g}^{(i)} \pmod{q} \quad : \{\textbf{v}_{i}\}_{i\in C_k}\}
\end{split}
\end{equation*}
\noindent, where $\textbf{u}_{ij}$ is the random mask shared between client $i$ and $j$, $\overset{\$}{\leftarrow}$ donates uniformly sampling from some field, and $\equiv$ denotes that the distributions are identical.
\label{lem:sec}
\end{lemma}

Lemma~\ref{lem:sec} illustrates that the distribution of updates with random
masks added is identical to uniformly sampling from $\mathbb{Z}_q^d$. Hence,
individual clients' updates are securely hidden inside the random masks added by
the secure aggregation, and a semi-honest server can infer zero information of
individual clients using only the aggregated updates.
In the following, we give the proof of Lemma~\ref{lem:sec} with induction on $n$, where $n$ is the size of clients set $C_k$, $n = |C_k|$.

\noindent \textbf{Base Case:}~When $n=2$, assume $C_k = \{i, j\}$, $i < j$, and $\sum_{i\in C_k}\textbf{g}^{(i)} \pmod{q} = c$, $c$ is a constant. The first elements of two distributions are $\textbf{g}^{(i)} + \textbf{u}_{ij} \pmod{q}$ and $\textbf{v}_{i}$ respectively, and they are both uniformly random sampled from $\mathbb{Z}_q^d$.
The second elements are $\textbf{g}^{(j)} + \textbf{u}_{ji} \pmod{q} = c - (\textbf{g}^{(i)} + \textbf{u}_{ij}) \pmod{q}$ and $\textbf{v}_{j} = c - \textbf{v}_{i} \pmod{q}$ respectively, which are the sum minus of the corresponding first elements.
Thus the distributions are identical.

\noindent \textbf{Inductive Hypothesis:}
When $n=k$, the lemma holds.

\noindent \textbf{Inductive Step:}
According to the inductive hypothesis, the left and right distributions of the first $k$ clients are indistinguishable.
We follow the protocol to generate the left transcript when the $(k+1)^{th}$ client is added to the shard.
To deal with the right-hand-side transcript, we first add the same randomness as the left-hand-side to the first $k$ updates and then subtract them from the total sum to get the $(k+1)^{th}$ update.
It is easy to prove that the first $k$ updates on the left and right follow the same uniformly random distribution and the $(k+1)^{th}$ update is the difference between the total sum and the sum of the first $k$ updates.
Hence, the left and right transcripts are indistinguishable.

\end{proof}

In case the readers are not familiar with the simulation proof technique, please refer to~\cite{Lindell2017} for more information.
\section{Proof of Corollary~\ref{cor:robust}}
\label{sec:proof-corollary2}

\noindent\textbf{Corollary~\ref{cor:robust}} (Robustness against Byzantine adversaries).
Given the number of clients $n$, the number of shards $p$, and the fraction of corrupted clients $\epsilon$, \system provides a mean estimation with dimension-free error as long as $12\epsilon n < p$.

\begin{proof}
In the following analysis, we assume that the updates from the shards follow an
i.i.d. distribution, and we have justified this assumption in \S~\ref{sec:iid}.

The fraction of malicious shards is bounded by the worst case where each malicious client is exclusively assigned to different shards: $\epsilon'\leq \frac{\epsilon n}{p}\leq \frac{1}{12}$.
Given the assumption above, we have satisfied all the requirements in Theorem~\ref{thm:filterl2}.
Hence, \system provides a mean estimation with dimension-free error as long as $12\epsilon n < p$. \qedhere

\end{proof}

\section{Proof of Corollary~\ref{cor:clt}}
\label{sec:proof-corollary3}


\noindent\textbf{Corollary~\ref{cor:clt}} (Ideally i.i.d. after Sharding).
Assume that the updates from the clients follow Definition~\ref{def:hetero} where $k \ll \frac{n}{p}$.
Besides, 
\begin{equation*}
\lim_{|H|\rightarrow \infty}\frac{1}{s_H^2}\sum_{i\in H}\mathbb{E}[(g_i-\mu_i)^2\cdot\mathbbm{1}\{|g_i-\mu_i|>\epsilon s_H\}]=0
\end{equation*}
, where $s_H^2=\sum_{i\in H}\sigma_{\phi(i)}^2$.
Given the uniform randomness of sharding, we can view the distribution index $\phi(j)$ as drawn from some distribution $\Phi$ on $[k]$.
Let $\bar{\mu}=\mathbb{E}_{i\sim\Phi}[\mu_i]=\sum_{x\in[k]}\Phi(x=i)\mu_i$ and $\bar{\sigma}^2=\mathbb{E}_{i\sim\Phi}[\sigma_i^2]=\sum_{x\in[k]}\Phi(x=i)\sigma_i^2$.
Then, 
\begin{equation*}
    \frac{1}{|H|}\sum_{i\in H}g_i\overset{d}{\rightarrow} N(\bar{\mu}, \frac{\bar{\sigma}^2}{|H|})
\end{equation*}

\begin{proof}
Given the assumption in Corollary~\ref{cor:clt}, we can apply Lindeberg CLT to the aggregated update distribution of a randomly selected shard.
We first re-organize Equation~\ref{eq:clt} to the following form: 
\begin{equation*}
\frac{1}{|H|}\sum_{i\in H}g_i\overset{d}{\rightarrow} N(\frac{1}{|H|}\sum_{i\in H}\mu_{\phi(i)}, \frac{s_H^2}{|H|^2})
\end{equation*}
Then we derive the mean and variance of the target distribution,
\begin{itemize}
    \item $\mathbb{E}[\frac{\sum_{i\in H}\mu_{\Phi(i)}}{|H|}]=\sum_{i\in[k]}\Phi(x=i)\mu_i=\bar{\mu}$,\\
    $\mathbb{V}[\frac{\sum_{i\in H}\mu_{\Phi(i)}}{|H|}]=\frac{\sum_{i\in[k]}\Phi(x=i)(\mu_i-\bar{\mu})^2}{|H|}=\frac{\sigma_{\mathbb{E}}^2}{|H|}$
    \item $\mathbb{E}[\frac{s_H^2}{|H|}]=\sum_{x\in[k]}\Phi(x=i)\sigma_i^2=\bar{\sigma}^2$,\\ $\mathbb{V}[\frac{s_H^2}{|H|}]=\frac{\sum_{x\in[k]}\Phi(x=i)(\sigma_i^2-\bar{\sigma}^2)^2}{|H|}=\frac{\sigma_{\mathbb{V}}^2}{|H|}$
\end{itemize}
As the two distributions have bounded variance, according to Chebyshev's inequality, 
\begin{equation*}
\lim_{|H|\rightarrow \infty }\frac{\sum_{i\in H}\mu_{\Phi(i)}}{|H|}=\bar{\mu}, \lim_{|H|\rightarrow \infty }\frac{s_H^2}{|H|}=\bar{\sigma}^2
\end{equation*}
Thus, when $|H|\rightarrow\infty, \frac{\sum_{i\in H}\mu_{\phi(i)}}{|H|}\rightarrow\bar{\mu}, \frac{s_H^2}{|H|^2}\rightarrow\frac{\bar{\sigma}^2}{|H|}$.
\end{proof}

\section{Empirical Validation of Corollary~\ref{cor:clt}}
\label{sec:emp-val}
\begin{wrapfigure}{r}{0.5\textwidth}
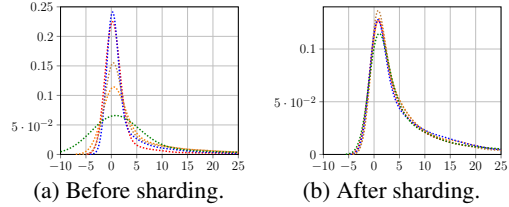

\captionsetup{font={footnotesize}}
    \vspace{-10pt}
     \centering
     \hfill
     \begin{subfigure}[b]{0.23\textwidth}
         \centering
         \resizebox{\textwidth}{!}{\input{figs/hetero/beforeshard}}
         \caption{Before sharding.}
         \label{fig:hetero_beforeshard}
     \end{subfigure}
     \hfill
     \begin{subfigure}[b]{0.23\textwidth}
         \centering
         \resizebox{\textwidth}{!}{\input{figs/hetero/aftershard}}
         \caption{After sharding.}
         \label{fig:hetero_aftershard}
     \end{subfigure}
    \hfill
    \caption{Distribution w/ (w/o) sharding. 
    }
    \label{fig:f3}
    \vspace{-10pt}
\end{wrapfigure}
Corollary~\ref{cor:clt} depicts the uniform convergence of the aggregated update distribution as the shard size grows.
To empirically validate our assumption, we simulate heterogeneous data
distribution by assigning MNIST samples with different labels to 25 clients.
These clients are split into five shards.
We plot the distributions of the updates before and after sharding as shown in Figure~\ref{fig:f3}.
Each line represents the weight distribution within one update.
Figure~\ref{fig:hetero_beforeshard} plots five updates from the same shard and Figure~\ref{fig:hetero_aftershard} plots the averaged updates from the five shards.
It is obvious that after sharding the distributions are more densely and identically distributed as discussed above.

\section{Details of Attack Methods Evaluated}
\label{sec:eval-attack}

The first and second attacks we used are the model poisoning attacks from~\cite{fang2019local}.
The model poisoning attacks aim to increase the error rate of the converged model even facing Byzantine-robust protocols.
In these attacks, the malicious clients search for poisoning updates by solving an optimization problem.
We employ two attacks proposed in their work targeting at Krum and Trimmed Mean. 
These two attacks are referred to as Krum Attack (KA) and Trimmed Mean Attack (TMA).

The third attack we considered is a backdoor attack from~\cite{bhagoji2019analyzing}.
The attack aims to insert a backdoor functionality while preserving high accuracy on the validation set.
Similarly, the search for the attack gradient is formalized as an optimization
problem and the authors tweak the objective function with some stealth metrics
to make the attack gradient hard to detect.
We refer to the attack as Model Poisoning Attack (MPA).

The fourth attack is also a backdoor attack proposed by~\cite{bagdasaryan2020backdoor}.
In this attack, the adversary locally trains a model with a backdoor and attempts to replace the global model with the local model by uploading the difference between the target model and the global model.
We refer to the attack as Model Replacement Attack (MRA) in the rest of the section.

The fifth attack is a distributed backdoor attack (DBA) from~\cite{xie2019dba}, where the attacker controls several clients and manipulates their updates to collaboratively insert a backdoor into the global model.

Attentive readers might notice that the first and second attacks are
specifically designed for Krum and TrimmedMean.
Careful readers may wonder about the feasibility of designing an attack particularly
for \agg.
We argue that it is non-trivial to design such an attack using the same idea
from~\cite{fang2019local} since the optimization problem becomes intractable
when \agg is plugged in.
Due to the theoretically stronger robustness, we assume that it is very
challenging to design targeted attacks for \agg like Krum or Trimmed Mean; we
leave it as one important future direction to design such an attack or
rigorously prove the impossibility.

\section{Details of Evaluation Setup}
\label{sec:eval-setup-append}
This section reports detailed information regarding model architecture,
hyper-parameters, and datasets processing procedures.

The model without attack and under KA, TMA, MPA, and MRA attacks is constructed
by two convolutional layers following two fully connected layers with ReLU as
the activation function.
Nevertheless, for the DBA attack, it cannot be successfully launched in simple
models, because simple models are likely to overfit the injected backdoors.
Therefore, we use the default model architecture which is ResNet18 evaluated in
the DBA paper~\cite{xie2019dba}.
 
For DBA, we use the default parameters used in the original
paper~\cite{xie2019dba} and use the multiple-shot attack strategy proposed by
that paper to evaluate the attack success rate under all estimators. As for the
other experiments, we use a learning rate $1e-3$, batch size $10$, and we
configure the same initial model state. We run $30$ epochs for these
experiments.
For estimator Trimmed Mean, we set the threshold as $30\%$, which means that it
can rule out $30\%$ of out-of-distribution updates.
For Bulyan that can tolerate $f$ Byzantine workers, we set $f=4$ which needs to
satisfy the assumption $n \geq 4f + 3$ required by Bulyan, where $n$ is the
number of clients.
For \agg, we set $\sigma = 1e-6$ and $\eta = 20$.
All other settings like random seeds are identical for all attack methods,
estimators, and training process to make the comparison fair.

The datasets we evaluate are MNIST and FashionMNIST, and we use both training
and testing splits provided by Pytorch.
The license of FashionMNIST is MIT license.
For the i.i.d. setting, the data is randomly distributed to all clients and each
client holds the same number of data samples.
For the non-i.i.d. setting, the data is distributed randomly to all clients and
each client holds the same number of samples with three different labels. The
labels are assigned to clients randomly.
We emphasize that for each experiment, we use the same random seed to split or
distribute the dataset to launch a fair comparison.

All the experiments were conducted on a Ubuntu16.04 LTS server with eight
Geforce GTX 1080 Ti. Please refer to our released codebase provided in \sm\ for
further implementation details.

\section{Evaluation Results Under Non-i.i.d. Settings}
\label{sec:eval-non-iid}

\begin{figure*}[htbp]
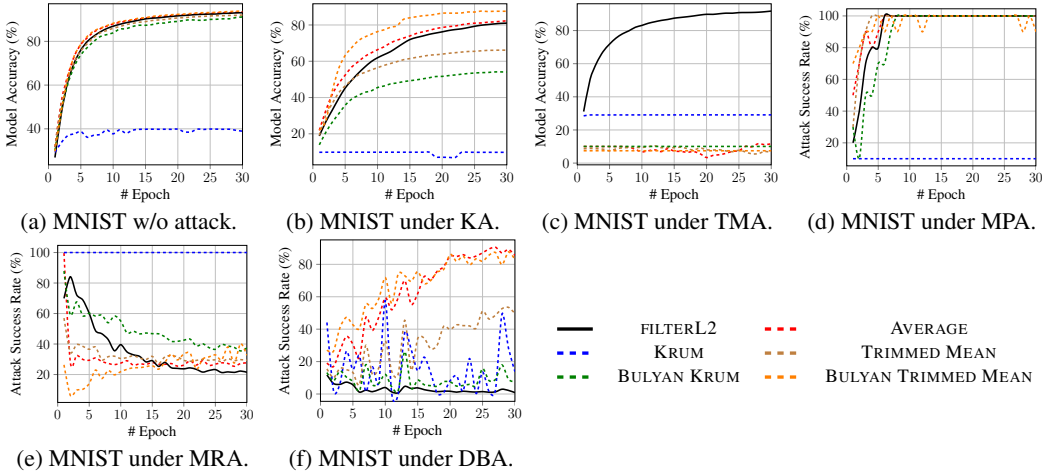

     \centering
     \hfill
     \begin{subfigure}[b]{\figwidth}
         \centering
         \resizebox{\textwidth}{!}{\input{figs/hetero/orig/INFIMNIST/withoutattack}}
         \caption{MNIST w/o attack.}
         \label{fig:mnist_noattack3}
     \end{subfigure}
     \hfill
     \begin{subfigure}[b]{\figwidth}
         \centering
         \resizebox{\textwidth}{!}{\input{figs/hetero/orig/INFIMNIST/krum}}
         \caption{MNIST under KA.}
         \label{fig:mnist_krum3}
     \end{subfigure}
     \hfill
     \begin{subfigure}[b]{\figwidth}
         \centering
         \resizebox{\textwidth}{!}{\input{figs/hetero/orig/INFIMNIST/trimmedmean}}
         \caption{MNIST under TMA.}
         \label{fig:mnist_trimmedmean3}
     \end{subfigure}
     \hfill
     \begin{subfigure}[b]{\figwidth}
         \centering
         \resizebox{\textwidth}{!}{\input{figs/hetero/orig/INFIMNIST/modelpoisoning}}
         \caption{MNIST under MPA.}
         \label{fig:mnist_modelpoisoning3}
     \end{subfigure}
     
     \hfill
     \begin{subfigure}[b]{\figwidth}
         \centering
         \resizebox{\textwidth}{!}{\input{figs/hetero/orig/INFIMNIST/backdoor}}
         \caption{MNIST under MRA.}
         \label{fig:mnist_backdoor3}
     \end{subfigure}
     \hfill
     \begin{subfigure}[b]{\figwidth}
         \centering
         \resizebox{\textwidth}{!}{\input{figs/hetero/orig/INFIMNIST/dba}}
         \caption{MNIST under DBA.}
         \label{fig:mnist_dba3}
     \end{subfigure}
     \hfill
    \begin{subfigure}[b]{0.48\textwidth}
        \centering
        \raisebox{\height}{\resizebox{\textwidth}{!}{\input{figs/main/non-iid-legend}}}
    \end{subfigure}
    \caption{Attack performance under different estimators with non-i.i.d. data.}
    \label{fig:f4}
\end{figure*}

In this section, we provide the evaluation result of \agg and \system in
non-i.i.d. settings. The performance of the six evaluated aggregators under
different attacks on MNIST is shown in \F~\ref{fig:f4}.
As we report in \S~\ref{sec:eval-results}, most of the estimators do not perform
well in the non-i.i.d. setting.
The i.i.d. assumption of the benign updates is broken, which makes it difficult
for the estimators to distinguish between benign and malicious updates.
However, \agg still achieves 91.75\% accuracy under TMA and reduces the attack
success rate to 21.46\% and 0.86\% at the end of the training process under MRA
and DBA attacks, respectively.
On the other hand, the accuracy drops to 81.01\% under KA, and the attack success
rate increases to 100.00\% under MPA. This is not surprising since the i.i.d.
assumption on which \agg is based is broken.

\captionsetup{font={footnotesize}}
\captionsetup[sub]{font={footnotesize}}

\captionsetup{font={footnotesize}}
\captionsetup[sub]{font={footnotesize}}
\begin{figure*}[htbp]
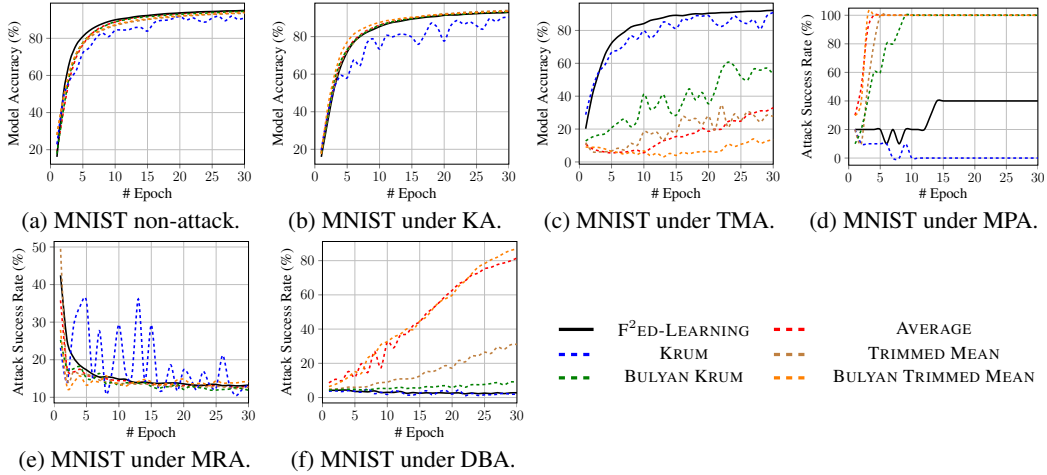

     \centering
     \hfill
     \begin{subfigure}[b]{\figwidth}
         \centering
         \resizebox{\textwidth}{!}{\input{figs/hetero/shard/INFIMNIST/withoutattack}}
         \caption{MNIST non-attack.}
         \label{fig:mnist_noattack4}
     \end{subfigure}
     \hfill
     \begin{subfigure}[b]{\figwidth}
         \centering
         \resizebox{\textwidth}{!}{\input{figs/hetero/shard/INFIMNIST/krum}}
         \caption{MNIST under KA.}
         \label{fig:mnist_krum4}
     \end{subfigure}
     \hfill
     \begin{subfigure}[b]{\figwidth}
         \centering
         \resizebox{\textwidth}{!}{\input{figs/hetero/shard/INFIMNIST/trimmedmean}}
         \caption{MNIST under TMA.}
         \label{fig:mnist_trimmedmean4}
     \end{subfigure}
     \hfill
     \begin{subfigure}[b]{\figwidth}
         \centering
         \resizebox{\textwidth}{!}{\input{figs/hetero/shard/INFIMNIST/modelpoisoning}}
         \caption{MNIST under MPA.}
         \label{fig:mnist_modelpoisoning4}
     \end{subfigure}
     
     \hfill
     \begin{subfigure}[b]{\figwidth}
         \centering
         \resizebox{\textwidth}{!}{\input{figs/hetero/shard/INFIMNIST/backdoor}}
         \caption{MNIST under MRA.}
         \label{fig:mnist_backdoor4}
     \end{subfigure}
     \hfill
     \begin{subfigure}[b]{\figwidth}
         \centering
         \resizebox{\textwidth}{!}{\input{figs/hetero/shard/INFIMNIST/dba}}
         \caption{MNIST under DBA.}
         \label{fig:mnist_dba4}
     \end{subfigure}
     \hfill
    \begin{subfigure}[b]{0.48\textwidth}
        \centering
        \raisebox{\height}{\resizebox{\textwidth}{!}{\input{figs/main/non-iid-legend2}}}
    \end{subfigure}
    \caption{Attack performance under different estimators with sharded secure aggregation and non-i.i.d. data.}
    \label{fig:f5}
\end{figure*}

The performance of the six evaluated aggregators with sharding under different
attacks on MNIST is shown in \F~\ref{fig:f5}.
As we illustrate in \S~\ref{sec:eval-results}, \system achieves optimal or
close-to-optimal performance among all the estimators in non-i.i.d. settings.
\system achieves a notably high 92.96\% accuracy, manifesting 11.95\%
improvement compared with \agg when no sharding is applied.
\system also reduces the attack success rate under MPA to 40.00\% compared to
without sharding which is 100.00\%.
This result further empirically validates Corollary~\ref{cor:clt} such that
sharding re-establishes the i.i.d. nature of the collected updates which is
necessary for \agg to be functional, as discussed in \S~\ref{sec:iid}.

An interesting phenomenon is that KA can be successfully defended by all
aggregators when the clients are
sharded~(\F~\ref{fig:mnist_krum2},\ref{fig:fashionmnist_krum2},\ref{fig:mnist_krum4}).
The reason is that KA is targeted at Krum without sharding and wants to maximize the probability that a malicious update is chosen by Krum.
Once integrated with sharding, Krum can only select from the averaged updates
provided by shards, and therefore, the effect of the malicious update is diluted.
This demonstrates that sharding itself can defend against some attacks by
diluting the effect of malicious updates.
Overall, we interpret the evaluation results as highly promising: benefitting
from the approximate i.i.d. distribution among shards, all aggregators perform
notably much better under different attacks compared with the non-i.i.d. setting
without sharding (\F~\ref{fig:f4}).

\section{Influence of Shard Size}
\label{sec:eval-shardsize}

\begin{wrapfigure}{r}{0.35\textwidth}
     \centering
         \resizebox{0.35\textwidth}{!}{%
            \input{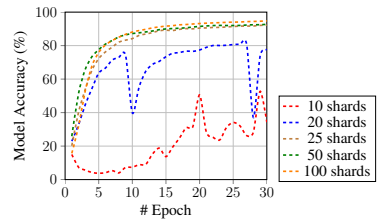}%
        }
    \caption{MNIST under TMA with various number of shards.}
    \label{fig:f6}
\end{wrapfigure}

\system introduces a new hyper-parameter, the number of shards $p$, which
specifies to what extent clients are divided into small groups. In this section,
we empirically evaluate the influence of this hyper-parameter on \system. We
report the results in \F~\ref{fig:f6}, where we launch the TMA attack toward
\system configured with different $p$.

We interpret that \F~\ref{fig:f6} illustrates the trade-off between security and
robustness guarantee via tuning $p$. When the number of shards equals the number
of clients, the system is trivially equivalent to FilterL2 without sharding and
can achieve optimal model accuracy. However, this setting sacrifices security,
given that the semi-honest server has access to each client's individual update
and \system provides no further security guarantee compared with vanilla FL. On
the other end of the spectrum, when the number of shards converges to one, the
system degrades to simply averaging each client's update; this extreme
configuration thus provides the strongest security but the weakest robustness.
When the number of shards falls within these two extremes, the model accuracy
gradually changes under the TMA attack, as depicted in \F~\ref{fig:f6}.
Overall, our empirical observation shows that $p=\frac{n}{4}$ ($n$ is the total
number of clients) would be a desirable choice when using datasets like MNIST
and FashionMNIST in our evaluation setting. Holistically, the optimal choice of $p$ should depend on the total number of clients, the requirement of security and robustness level, and
also the specific FL task.











\end{document}